\documentstyle[epsfig]{aipproc}

\def\thesection{\Roman{section}.}
\def\thesubsection{\arabic{subsection}.}               

\begin{document}
\title{Intersections 2000: \\[.8ex] What's New in Hadron Physics
\thanks{Work supported by the Department of Energy under contract
number DE--AC03--76SF00515.}}

\author{James D. Bjorken}
\address{Stanford Linear Accelerator Center\\
Stanford University, Stanford, California 95409}

\maketitle

\begin{abstract}
Hadron physics is that part of QCD dealing with hadron structure and
vacuum structure, almost all of which is nonperturbative in nature. Some
of the open problems in this field are outlined. We argue that hadron
physics is a distinct subfield, no longer within particle physics, and not
at all the same as classical nuclear physics. We believe that it needs to
be better organized, and that a first step in doing so might be to
establish hadron physics as a new division within the American Physical
Society.
\end{abstract}

\section{The Big Picture}

The main portion of this talk deals with the subject of hadron physics:
what it is, what some of its challenges are, and why I believe the
hadron-physics community needs to identify itself more strongly and
precisely in order to define and protect its long range experimental
program. But before turning to that, it may be of use to put this subject
in the context of the bigger picture of basic particle-physics goals.
Another reason is that this conference has not been just about hadron
physics. By my count about 60 percent of the parallel sessions dealt with
hadron-physics issues, while only 45 percent of the plenary talks were on
hadron physics, the remainder dealing with the bigger picture.

For most of the last twenty years, the Big Picture in particle physics has
centered around the Big Three issues, namely Higgs, SUSY, and CP. I
believe, as does most everyone, that the most important of these is the
problem of mass and the nature of the Higgs sector, responsible for
electroweak symmetry breaking. While this question seems timeless, having
been around in an almost unchanged form for over two decades, our
perspective of it has actually shifted somewhat. Thanks to the discovery
at Fermilab of a very heavy top quark, and to the many beautiful precision
electroweak measurements from CERN, SLAC, and elsewhere, the mass of the
Higgs boson cannot be too large. This is encouragement that we will within a
decade have a direct experimental handle on the question, from experiments
at the Fermilab Tevatron and/or the CERN Large Hadron Collider.

How this turns out experimentally will have a profound influence on the
future of the field.  I like to contrast the options in terms of two
extremes. One is the ``desert" scenario, where the theory remains
essentially what we now have all the way up to a very high mass scale, for
example the Grand Unification scale of $10^{15}$ GeV or so.  This can only
occur if the Higgs boson, the only undiscovered particle that the desert
scenario requires, has a mass of 160 $\pm$ 20 GeV. Other masses are ruled out
by the requirements of vacuum stability and the absence of strong Higgs
self-interactions (cf. Fig. 1). The other extreme is that of the
supersymmetric extension of the standard model (SUSY, or MSSM), where each
known particle has its superpartner, differing in spin by one-half unit,
and with the superpartner masses typically less than a TeV. Also, the MSSM
Higgs sector is larger, with at least one member expected to have mass
less than about 130 GeV.

\begin{figure}[htb] 
\centerline{\epsfig{file=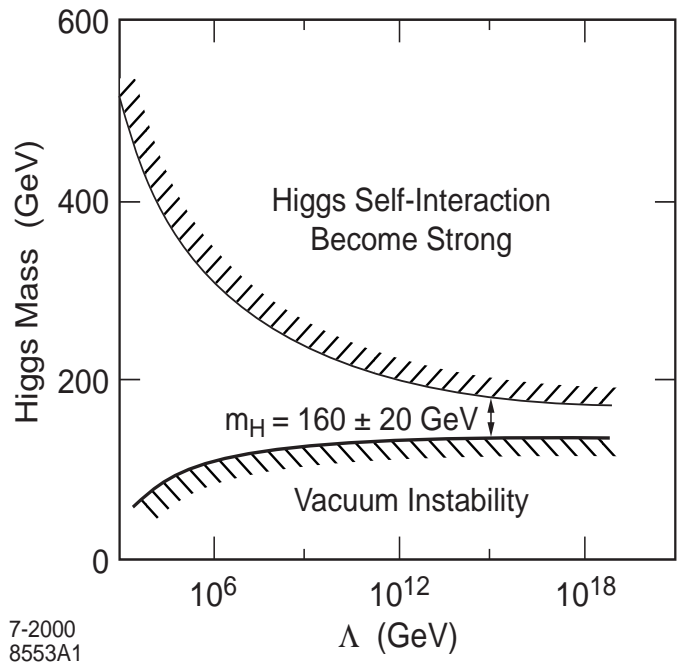,height=3.5in,width=3.5in}}
\vspace{10pt}
\caption[]{Values of $m_{\rm Higgs}$ and momentum scale for which the
Standard Model exists, {\it i.e.} where electroweak perturbation theory
converges.  The upper region is forbidden because the self-interactions
of the Higgs particle become strong.  The lower region is forbidden because
the vacuum itself becomes unstable.}
\label{fig1}
\end{figure}

If the SUSY scenario is correct, there will be full employment for
experimentalists. They not only need to discover the superpartners, but
also to determine the more than 100 extra fundamental parameters that
characterize this extension of the standard model. On the other hand, if
only the 160 GeV Higgs boson is seen, and nothing else new is found, this
will be rather strong direct evidence for the desert scenario. In that
extreme, there would be no reliable, landmark, higher-mass scales for new
experimental facilities to aim for. It would become more difficult to
justify multi-billion dollar future colliders were one to be unable to
certify in advance new discoveries, in the way that has been done for
previous facilities ($W$ at the SPS, top at the TeVatron, Higgs at the SSC
and LHC). For these reasons alone, I see the outcome of the Higgs search
as a crucial turning point for the future of the field.

The desert scenario is rather unpopular because of the hierarchy problem,
namely the problem of why the Higgs mass remains so low when the natural
scale for it, via quadratically divergent radiative corrections, appears
to be much larger. This has led to the demand that something be invented
to cure the problem, the leading candidate being the MSSM. However, the
cosmological constant suffers from a very similar situation, one for which
a straightforward application of SUSY does not work. So it seems to me
that a serious, viable approach to the Higgs hierarchy problem is to
ignore it for the present, arguing that a much deeper source of the
solution, at the level of what is required for the cosmological constant
problem, is required. And once the hierarchy problem is ignored, the
``desert" theory is really very consistent, with no trace of the quadratic
divergence problem remaining, after renormalization, in the phenomenology.

While the Higgs situation has not changed all that much in the last two
decades, this does not mean that the Standard Model has remained
unchanged. Thanks to the strong evidence that neutrino oscillations really
exist, we now have a New Standard Model to replace the venerable Old
Standard Model. Instead of the twenty or so parameters characterizing the
Old Standard Model, we now have thirty or so parameters for the New
Standard Model, the exact number depending a bit on what one wants to
include in the count. For sure there are three neutrino masses and four
CKM-like mixing angles to determine. In addition perhaps the masses of the
three heavy Majorana particles of the seesaw mechanism should be included
as parameters, as well as a couple of phase factors seen at best only in
double-beta-decay processes and the like.

There is, in addition to the new parameter count, a definite shift at the
GUT level from $SU(5)$-like thinking to $SO(10)$-like thinking. All this to me
represents a significant advance, despite the presence of the extra
parameters that require explanation. As evidenced in this meeting, there
clearly will be increasing emphasis on the neutrino sector in the future:
it carries with it more than 30 percent of the parameters of the New
Standard Model, and these parameters will be at least as difficult to
determine as accurately as the CKM parameters, at present the focus of the
$B$-physics program. But twenty years ago, the determination of CKM phases
seemed to be a remote experimental possibility. Hopefully future progress
on the neutrino front will parallel what is happening now in the realm of
$B$ physics.

\section{Hadron Physics: What is It?}

The main thrust of this talk has to do with hadron physics.  I define it
as the physics of hadron structure and of (strong-interaction) vacuum
structure.  This puts it as a subfield of quantum chromodynamics (QCD),
just as chemistry, condensed-matter physics, and atomic physics are
subfields of quantum electrodynamics (QED).

In their infancy those three subfields were part of elementary particle
physics, but now are not. This was also the case for nuclear physics. And
I think the same has already happened to hadron physics. Most elementary
particle physicists, including those in positions of influence, do not pay
much attention to the issues in this field \cite{ref1}. And in fact most of the
experimental research in this field is done within the nuclear physics
community, even though it is a stretch to identify hadron physics with
nuclear physics. Because experimental (and of course theoretical) hadron
physics research spans all energy scales, this creates social and
organizational problems, a subject I will return to later \cite{ref2}.

To deal with these social issues I think that it is fundamental and
important to define in detail what hadron physics encompasses, what its
long range goals are, and what experimental and theoretical programs are
necessary to attain those goals. I cannot by myself articulate this here
in full. Hadron physics is very a big subject and I am sure to make errors
of omission, and to bias the subject matter toward my own particular
interests. In fact a better method might be simply to peruse the contents
of the proceedings of this conference. In the next section, I will simply
catalog some open problems I find interesting, as examples of the huge
challenges that are present in this subject, challenges which exist at
quite fundamental levels. And because the subject matter of hadron physics
rarely allows reliable perturbation-theory calculations, real progress
requires a data-driven approach, characterized by close interaction
between theory and experiment. In the final section I will return to the
social issues confronting hadron physics.

\section{Some Open Problems in Hadron Physics}

QCD, the basic theory of the strong interactions, is at short distances a
perturbation theory of the pointlike quark and gluon constituents of
hadrons. At very large distances QCD is a theory of pions and nucleons
(and their strange counterparts), and is characterized by spontaneous
symmetry breaking of the approximate chiral symmetry of QCD. At
intermediate distance scales, there is the rich arena of {\it e.g.} hadron
resonances, Regge trajectories, soft diffraction, and hadronization of the
partons, just to name a few of many topics.

But the physics at all distance scales is linked, and it is hard to find a
situation, even within the relatively clean, perturbative regime, where
the nonperturbative effects do not enter. Our first example is chosen to
illustrate this phenomenon.

\subsection{Perturbing the Chiral Vacuum}

A classic way of trying to understand the properties of a macroscopic
system is to perturb it with a small, localized impurity and study its
response. In the case of the chiral vacuum of long-distance QCD, a nice
way of doing this is by putting a small color dipole, such as heavy onium,
into the vacuum and examining its response. This implies creation at large
distances of a very weak pion cloud around the onium. The importance of
this cloud can be assessed by putting in another small dipole, and
determining the long range force between them, due to essentially two-pion
meson exchange. This has been done elegantly and cleanly by Fujii and
Kharzeev \cite{ref3}, who find that this force dominates at separations greater than
about 0.6 fermi. To be sure the potential energy associated with this
effect is quite small, under 1 MeV.

Nevertheless, this effect can be amplified by putting the two dipoles into
motion. At the qualitative level, one can see that the original clouds
will be compressed into pancakes when the onia become
extreme-relativistic. And the original rest energy of a pion cloud,
however small, can be turned into an arbitrarily large amount of
energy-momentum of a pionic pancake if an arbitrarily large boost is
applied. Now boost the two onia in opposite directions and put them into
collision, again at a large impact parameter. When the momentum density in
each of the pancakes exceeds, say 1 GeV/fermi$^2$ in the overlap region,
there will be ample amounts of cms energy available for particle
production, and the two onia should act like light hadrons as far as their
collision properties are concerned: in the jargon of the trade they will
``exchange a soft Pomeron".

Now the study of onium-onium collisions at extremely high energies is a
favorite playground of perturbative-QCD theorists. This is the so called
BFKL regime, where much effort has gone into summing up Feynman diagrams
to obtain a candidate phenomenology of ``hard Pomeron exchange" \cite{ref4}.
Nevertheless, the above argument implies that, for a fixed size of the
dipoles, however small, if one goes to high enough energy the perturbation
theory approach is destined to fail, and the soft physics is destined to
re-emerge. The smaller the dipoles are, the higher will be the energy
scale at which this phenomenon occurs.

The pion-cloud argument is not inconsistent with BFKL ideology, which also
anticipates a similar phenomenon occurring, due to ``diffusion of gluon
ladder transverse momenta into the infrared". However, I am not sure that
the energy scale where the transition occurs is the same in the two
approaches. But the bottom line remains the same: no matter how hard one
works for cleanliness in short-distance QCD, the soft physics usually
finds its way into the picture.

\subsection{Foundations of Perturbative QCD}

Perturbative QCD (pQCD) is a highly sophisticated and well developed
subject. However, at a fundamental level there are, I believe, some real
problems. The basic issue is that, despite the fact that it is commonly
done, it is not legal in pQCD to put the quarks and gluons on mass shell,
{\it i.e.} to treat them as asymptotic states that propagate to infinity. They
clearly do not---there is no $S$-matrix for quarks and gluons. It is rather
ironic that in the old days before QCD one had an $S$ matrix formalism
without a field theory, while now we need a field theory formalism without
an $S$-matrix.

To see the problem more concretely, it is only necessary to look at the
classic process of electron-positron annihilation into hadrons, in lowest
order. The associated vacuum polarization amplitude is a quark loop (Fig.
2) which at large spacelike $Q^2$ is a safe short-distance calculation. Its
absorptive part at timelike $Q^2$ is essentially the cross section of
interest. It is not completely ultraviolet safe, requiring an energy
average to make it safe. But even with that done, our problem emerges when
one wants to look at the angular distribution of the quark-antiquark
dijets which build the total cross section. To get that differential cross
section, one typically calculates the absorptive part as if the quarks
could be put on mass shell---which \break
we must admit is illegal.

\begin{figure}[htb] 
\centerline{\epsfig{file=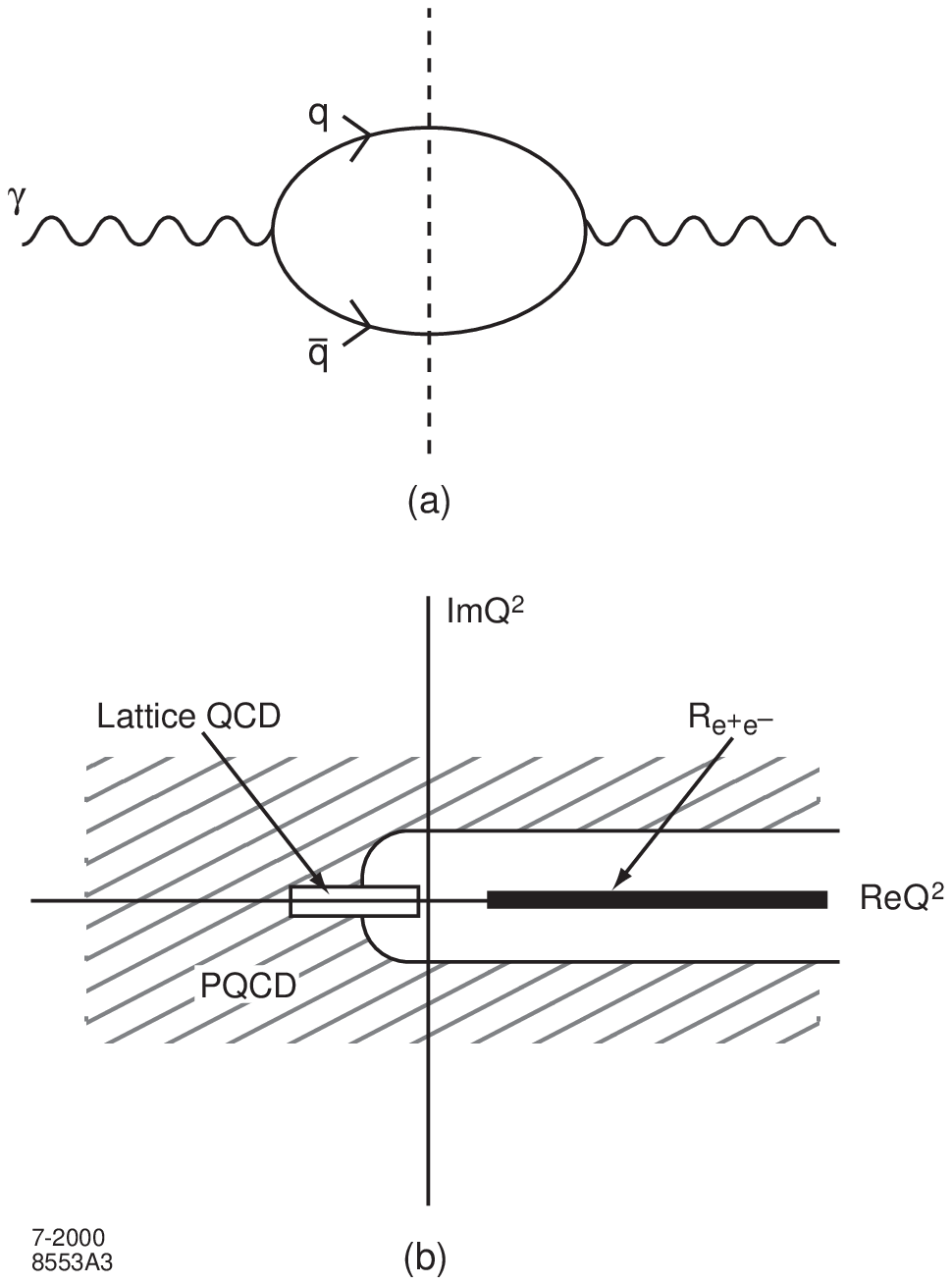,height=3.5in,width=3.5in}}
\vspace{10pt}
\caption[]{(a) The vacuum polarization amplitude whose absorptive part
describes the quark-antiquark dijet final state in electron-positron annihilation.
(b)  The complex $Q^2$ plane appropriate to this process; only in the shaded
region is perturbative QCD justifiable.}
\label{fig2}
\end{figure}

That this is not academic can be seen by imagining a QCD where only the
bottom quark exists. With no light quarks available, the $b$ and $\bar b$
quarks will (probably) be connected by an essentially unbreakable QCD
string or flux tube. This leads (probably) to the conclusion that the
final states will be a dense spectrum of excited onia, with no jets to be
seen.  The only way jets could occur is through glueball emission, and
this appears to be a highly inefficient mechanism.

So the bottom line is that identification of final state jets with
on-shell partons is a model assumption, which at present lacks a firm
foundation. To be sure, it is an eminently reasonable assumption. But it
would be better to have a sounder line of argument.

\subsection{Parton Correlations and Multiplicities}

How many quarks are there in a proton? ``Three", says the spectroscopist.
But the deep-inelastic community will (or should) answer ``infinity". Both
answers have their place, but connecting the two is still a problem. For
example, there is a substantial sub-community of hadron physicists, in
particular the practitioners of ``exclusive QCD", who use a Fock-space
description of the partons comprising a light-cone proton, with the
``leading Fock-space component" having three and only three quarks in it.
Now given that the average number of quarks is infinite (being essentially
proportional to the integral over $\ell n\, x$ of $F_2(x)$), this would mean that
the multiplicity distribution of partons is quite peculiar, with the
``Fock-space" piece of it of finite mean multiplicity (with how much total
weight, please?), while the rest is of infinite multiplicity.

I am simply baffled that this inconsistency of approach seems not to be
recognized at all as a problem.  When I mention it to others, the response
seems to be that I am the one with the problem. Maybe this is so. But
maybe there is a clue in the example of pion clouds around onium in item
1. A partonic description of the collision process described there
(especially if one of the onia is replaced by a spacelike photon, which
also makes a splendid small color dipole) leads essentially to a parton
distribution as shown in Fig. 3. The region to the right is essentially
perturbative and probably amenable to the Fock-space, perturbative
methodology. But present in addition are all the nonperturbative partons
comprising the pion cloud.  For phenomenology which concentrates on the
large-$x$ valence system, the cloud partons are presumably inconsequential.
However, as one goes from the case of small color dipoles to realistic
light hadrons, the role of the cloud partons becomes much more important,
and in the light-quark limit one must work hard to justify their neglect.

\begin{figure}[htb] 
\centerline{\epsfig{file=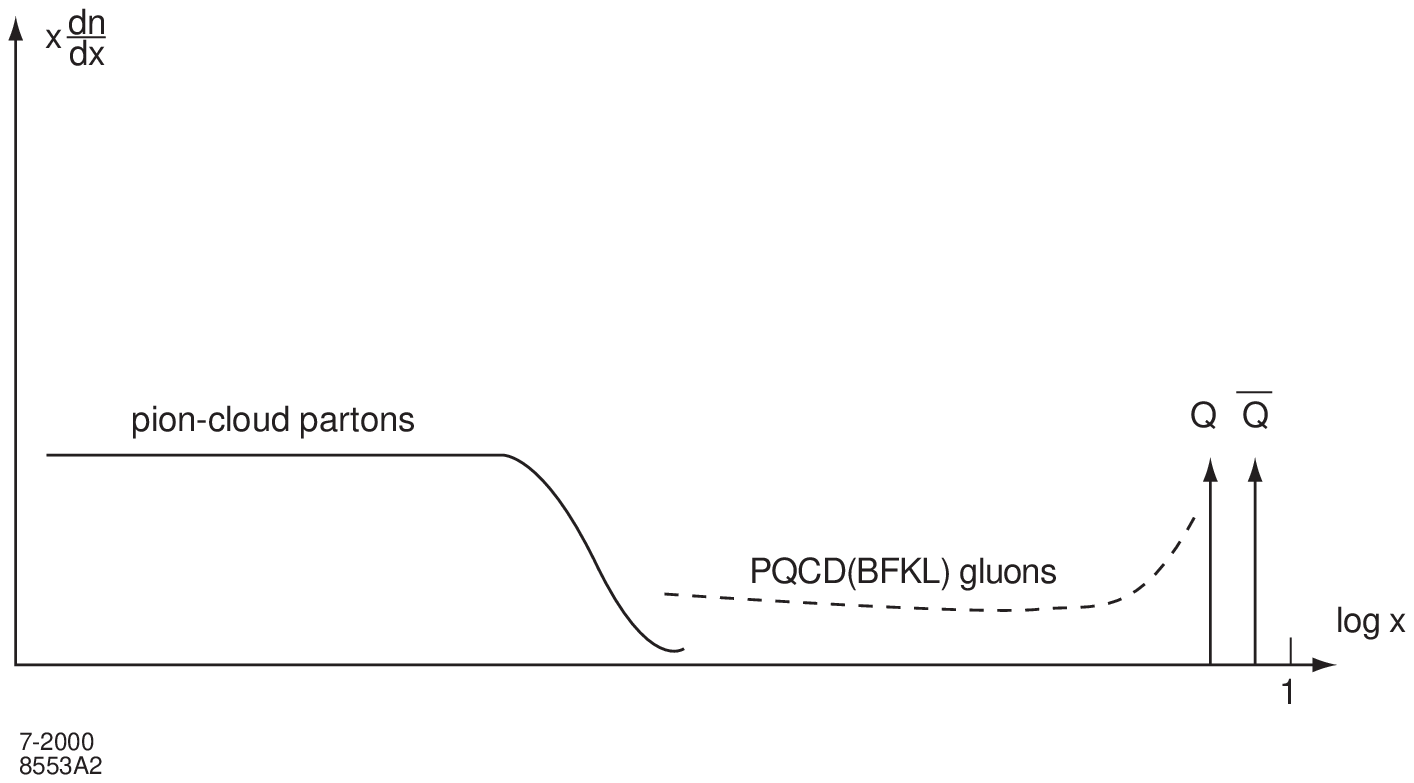,height=2.5in,width=5in}}
\vspace{10pt}
\caption[]{The parton distribution expected for a small, massive color dipole.
The BFKL region where perturbative QCD may be applicable is of width
$\sim\ell n\, Q^2d^2$, where $d$ is the (color) dipole moment of the
onium.  A similar picture should apply for the structure function of a virtual
photon with squared momentum $P^2 \sim d^{-2}$.}
\label{fig3}
\end{figure}

Even leaving this issue aside, the parton multiplicity distribution itself
is poorly understood, to say the least. Is it Poissonian, or KNO? How
might one distinguish one from the other? And the correlations of the
partons in the transverse plane are largely unknown. For example, are most
of the infinite sea of wee partons inside the three constituent quarks,
are they mostly outside, or are they mostly uncorrelated?

There are good reasons why such simple questions remain unanswered, and
most of them have to do with the fact that it is very hard to get at them
experimentally.  Double parton---or multiple parton---collisions have the
potential to provide information on the correlations. While some
experimental work has been done already, this subject will become
increasingly practical at the LHC \cite{ref5}, provided that at least one of the
several thousands of experimentalists working there will care enough to
make the measurements and to do the analysis.  Nevertheless, even in the
absence of data, it still might be interesting to have the various
theoretical options compete with each other at the Monte-Carlo simulation
level, in order to search for sensitive experimental indicators.

\subsection{Spectroscopy of Light Quarks}

Thus far, our examples have swung from the very large distance regime to
the very short distance regime, with minimal emphasis on the intermediate
distance scale. That scale is the richest phenomenologically, and is
certainly the crux region to understand, in order to obtain a
command of what QCD is really about. And at the heart of the subject is
the hadron spectrum, in particular the spectrum of hadrons built from
light quarks. There is a long and distinguished history of hadron
spectroscopy. It deserves at least half, and probably more than half, of
the credit for the establishment of the standard-model quark picture of
hadron structure. For me, a high point of hadron spectroscopy occurred in
the mid 1970s with the very sophisticated measurements and analyses of
baryon resonance spectra. An enormous body of work could be summarized by
the SU(6) classification ``{\bf 56}, $L$ even; {\bf 70}, $L$ odd", very consistent with a
quark-diquark picture of baryon structure \cite{ref6}. I am not sure how well this
picture has survived the subsequent 25 years.  But whatever the present
situation is for baryons, there has never been such an easy summary of the
situation regarding the meson multiplets. For a given choice of $J^{PC}$, the
lowest lying multiplet may be in good shape. But as soon as one looks at
higher excitations, there are missing states and there are extra
states, as we heard at this meeting from Jim Napolitano \cite{ref7}. Without question,
there is a great need---and opportunity---for a new round of experiments,
especially utilizing hadron beams.

At present multiparticle spectrometers such as MPS and Omega are being
phased out, and the only replacement for the future is a new facility in a
photon beam at Jlab. From the technical point of view, it seems to me that
it should be possible with state-of-the-art technology (as used by {\it e.g.}
ATLAS, CMS, CDF, D0, {\it etc.}) to create an ``electronic bubble chamber" for
hadron-induced processes with acceptance, resolution, and particle
identification capability at least as good as the old bubble chambers, and
with rate capability better by a millionfold. And the analysis power for
partial wave analyses or amplitude analyses likewise should exceed what
was done then by a millionfold. So a next generation attack across the
board would seem to me to be natural and to be potentially extremely
productive. It should be the case that a command of hadron resonance
spectra up to and beyond a mass scale of 2 GeV should yield a great deal
of understanding of the systematics of hadron structure.

\subsection{Non-Singlet Regge Trajectories}

One of the most profound and fertile topics in pre-QCD strong interactions
was that of the Regge theory of complex angular momentum and of Regge
poles. It is a subfield with a distinguished heritage, leading, via
duality and the Veneziano amplitude, to the creation of string theory and
the modern superstring industry.

In the context of hadron physics, Regge theory remains of great
significance. Most importantly it works. The experimental evidence for the
existence of Regge trajectories is in some cases extremely quantitative,
in particular for the trajectories containing vector mesons. This is
especially impressive for the very pure power-law energy dependences of $K$
regeneration amplitudes, recently reviewed for engineering reasons by the
KTeV collaboration \cite{ref8}.

Specific properties of the Regge trajectories are strong indicators of
the underlying dynamics. In particular linearly rising trajectories
suggest strongly a QCD string type of dynamics. This again places emphasis
on spectroscopic measurements at high spin and at high mass scales. In any
case the Regge picture systematizes in an important way the properties of
the resonant states.

I think much more work could be done in this field. As best I can recall,
it was only in the mid-seventies when the Regge behavior became
well-established experimentally, more or less concurrent with the
discovery of the $\psi$ and the subsequent change in direction of the field
as a whole away from that line of research. So a return visit and
systematic study in hadron-hadron collisions, with modern detectors and
analysis techniques, might be very productive.
 
One area of special importance is that of non-singlet Regge behavior in
deep-inelastic processes. In some processes there is evidence for the
expected Regge behavior in the scaling limit, {\it e.g.} the neutron-proton
difference, in the $F_3$ for neutrino reactions, and in the integrand of the
Adler neutrino sum rule. However, for the spin sum rule that bears my
name, there seems to be evidence for slow convergence of the sum at small
$x$, while Regge arguments would suggest that rapid convergence should be
the case. Examination of the Regge limit of the GDH sum rule in
photoproduction and its extension to low $Q^2$ electroproduction could be of
considerable use. But aside from details, measurements and analyses at
small $x$ which incisively test for Regge asymptotics are rare in the
contemporary deep inelastic scattering culture. Given the enormous amount
of attention paid to deep inelastic scattering in general, I find this
situation perplexing.

And just at the theoretical level, does QCD imply that Regge-pole
contributions scale, or should they be of higher twist? Can the Regge
residues be calculated from pQCD or something close to it? And while there
is a great deal of attention (rightfully) paid by theorists to the vacuum
Reggeon singularity (Pomeron), the structure of non-singlet  Regge-pole trajectories
should if anything be an easier problem. I believe they deserve a closer
look by theorists.

\subsection{Heavy Quark Spectroscopy}

With the high statistics and superb quality of recent charm and bottom
physics experiments, the opportunities for incisive spectroscopic studies
have increased dramatically. Things have gotten to the point that the
statistics of decays such as $D \rightarrow K \pi \pi$, or even $D 
\rightarrow 3 \pi$  is so high
that examination of the Dalitz plot yields useful information on
spectroscopy of ordinary mesons made of light quarks \cite{ref9}.

But the heavy quark excitations are themselves very interesting. The onium
systems are so clean that pQCD is the most appropriate starting point.
And their final states are fertile territory for glueball searches.
Especially interesting to me are the $D$ and especially $B$ mesons, where the
machinery of heavy quark effective theory can be applied. What this boils
down to is that everything having to do with the heavy quark is relatively
trivial, computable within pQCD, leaving the nontrivial system something
very close to a single constituent quark. In fact, a viable definition of
a constituent quark is the $B$ meson ``without" the $b$ quark. So the
excitations of $B$'s quite directly probe the properties of the single
constituent quark: for example its couplings to pions and photons, its
mass, its size, and (if the energy scale of the $B^*$ excitations can be made
large enough) any intrinsic excitations of its own.

While the electron-positron $B$ factories are ill-suited for this kind of
physics, the hadron-hadron colliders are very well-suited.  And in
addition to the intrinsic hadron-physics interest in the classification
and study of $B^*$ and other excitations of bottom hadrons, there is a good
engineering reason to do so. $B^*$s can help distinguish a secondary $B_d$
from a $B_{\bar d}$ in CP studies, freeing the experimentalist from having to
find a second tagging $B$ in the event.

\subsection{Confinement, Instantons, and the Vacuum}

From the point of view of theory, the belief that QCD is a viable theory
at all distance scales, despite the intractability of perturbation theory
in the infrared, rests on a hypothesis---that of confinement. There is
pretty good evidence for this from lattice calculations, not to mention
the experimental facts---which include the fact of our existence.
Nevertheless, there is as yet no consensus amongst theorists as to the
mechanism that is responsible for confinement. This is probably the
leading outstanding problem in all of hadron physics.

The confinement problem is closely linked to the problem of vacuum
structure. We have already alluded to the presence of chiral symmetry
breaking, leading to a chiral condensate at large distance scales. In
addition to that structure, there is the vacuum structure induced by the
occurrence of gauge potentials with nontrivial topology, and the existence
of instanton-induced transitions between vacua with differing gauge
topologies.  This creates both good news (existence of mass of the $\eta^\prime$
meson) and bad news (the possibility of CP violation in the strong
interactions).

When the instantons were first discovered, their effects were in poor
theoretical control. But at present the situation seems to be much better.
The instanton size distribution seems to be sharply peaked about a value
characterized by a momentum scale of 600 MeV. The density (in Euclidean
space-time) is relatively low, so that the fraction of spacetime
containing instanton fields is only a couple of percent or so.  And there
is a rather convincing line of argument that this instanton population
distorts the Dirac sea of light quarks in just the right way to induce
chiral symmetry breaking. It would be nice if the argument were to go
further and account for confinement as well, but this seems to be much
less likely.

It would also be nice to have a better handle experimentally on
instanton-induced effects. Efforts have been made to search for signatures
of instanton effects in multiparticle final states in collision processes,
although this is very difficult and speculative territory \cite{ref10}. I have my own
favorite candidate for a ``smoking-gun"  instanton-induced effect, namely
the leading decay modes of the $0^- \,\eta_c$ charmonium state. They are $\eta \pi
\pi$, $\eta^\prime \pi \pi$, and $ \bar K K\pi$, each with about a 5\%\ branching ratio, and
each being a state naturally produced via the 't Hooft instanton-induced
interaction:
\begin{equation}
{\cal L}\sim (\bar cc)(\bar uu)(\bar dd)(\bar ss) \ .
\end{equation}
I think close theoretical and experimental attention to these modes might
well be useful.

\subsection{Equations of State and Quark-Gluon Plasma}

Macroscopic properties of QCD are described by equations of state, which
can be studied as a function of the parameters of the theory, including
quark masses, number of colors, and of course temperature and chemical
potentials. There is plenty of activity and theoretical progress, as
described here by Krishna Rajagopal \cite{ref11}. And of course the heavy ion program
provides plenty of experimental impetus. The future looks bright indeed.

I believe that the heavy-ion programs, especially from RHIC and the
LHC,  will have important spinoffs into
high energy physics in at least two respects. One is that if the goal of
observation of quark-gluon plasma is achieved quantitatively, measurement
of the critical temperature should provide a quite good, competitive value
of $\Lambda_{QCD}$. More generally, the methods by which ion-ion collisions are studied,
with the emphasis on space-time evolution and hydrodynamic flow, will be
of great use in dealing with generic hadron-hadron collisions at Tevatron
and especially LHC energies, where the number of parton-parton
interactions per collision rivals the number of nucleon-nucleon
interactions per Au-Au collision.

\subsection{High Parton Densities at Very High Energies}

The energetic proton carries with it a very large number of partons, in
particular wee gluons, when it has momentum of a TeV and above. This was
anticipated by pQCD theorists, and has been well established by the measurements at
HERA.  As we mentioned above, this has great implications for central,
generic collisions at LHC energies, where the phenomenology of typical
collisions is expected to differ sharply from that at low energies,
essentially because opaque discs of dense gluons are coming into
collision. Even at the partonic level there will be strong absorptive
effects, copious minijet production, and possibly collective flow. There
deserves to be at the LHC (as well as at the TeVatron) serious attention
paid to the commonplace collisions as well as the high priority rare ones \cite{ref12a}.
There is a frontier of new physics to be explored.

I cannot resist mentioning here a vaguely related, speculative application
of high gluon-density physics for RHIC. Consider a RHIC Au-Au collision,
not in their laboratory, cms frame, but in a frame where one of the Au
nuclei is at rest. What that Au nucleus sees is a very energetic ion
bearing down on it, carrying momentum of about 20 TeV per nucleon. There
clearly is an enormous wee-gluon density that the rest-frame ion sees.
With probability unity, each parton in each rest-frame nucleon gets hit by
a gluon, and is ``Compton-scattered" into a relativistic final state, with
large longitudinal laboratory momentum.

What this essentially means is that everything that was in the rest-frame
nuclear matter gets swept up by the projectile and carried away with it at
the speed of light---the nuclear matter is stuck to the pancake. So in the
volume originally occupied by the resting ion there is ``nothing" left.

While this ``nothing" is probably not vacuum, it cannot be highly excited
either, at least that part of the ``nothing"  which will radiate
secondaries more or less isotropically. This follows simply from
conservation of $E - p_z$. In the initial state the value of $E - p_z$ is
essentially $A$ GeV, and this is spread over a large volume. So the density
of $E - p_z$ in the final state must be low, no more than 140 MeV/fermi$^3$.
This would seem to imply that at most only soft pions can be isotropically emitted
from the ``nothing". Perhaps these pions could be coherently emitted, a la
Bose condensation or via a disoriented-chiral-condensate (DCC) mechanism.  
In the RHIC laboratory frame, this would imply a secondary ``beam" of these
pions emerging in the forward direction, with the same velocity as the
incident ion beam. In other words, given 100 GeV/nucleon incident
momentum, this cluster of pions would emerge with 14 GeV/pion, and with a
transverse-momentum spread perhaps as low as 100 MeV. Such pions would be
difficult to detect with the existing detectors, although it is not hard
to envisage detectors which could do the job \cite{ref13}.

\subsection{The Approach to Scaling}

A very large amount of activity now exists in electron-nucleon and
electron-ion scattering at intermediate energies, especially at Jlab,
where a variety of very beautiful measurements are emerging. Here the
basic challenge, as I see it, is to map out in detail the transition from
manifestly long-distance descriptions ({\it e.g.} elastic $e$-$p$ scattering at
relatively small $Q^2$) to manifestly short distance descriptions, as used
in the scaling region in deep-inelastic scattering. There are a variety of
kinematic regions to experimentally explore, the two most important ones
being high $Q^2$ at small inelasticity (``exclusive QCD") and low $Q^2$ at
large inelasticity (``soft Pomeron physics"). Both are merged at moderate
values of $Q^2$ and inelasticity with the physics of resonances and
non-singlet Regge dynamics.

Among the theoretical challenges is the search for the best descriptive
tools. Help comes from duality concepts, both of the Bloom-Gilman type as
well as the Regge-resonance type, not to mention parton-hadron duality in
its more general context. Sum rules are a powerful tool. While much more
can be done with such tools, I think the time has come to search further
for specific dynamical descriptions good enough to incorporate the sum
rules (in particular a fully relativistic description) and other general
features. An excellent example of what I have in mind is the chiral quark
model of hadrons, as laid out by D. Diakonov, M. Polyakov, and their
collaborators \cite {ref14}. They start at the fundamental level of short distance QCD,
integrate out instanton effects, and find a viable description of hadrons
in the chiral sector which still is consistent with relativity and deep
inelastic phenomenology. While the assumptions made are perhaps not under
complete theoretical control, the credibility level is still very high,
and capable in the future of going higher.

\subsection{Diffraction}

High energy hadrons, as extended objects, are nearly black discs. Their
elastic scattering amplitudes provide clear evidence that this is the
case. However, the shadows that hadrons cast in their high energy
interactions are much more interesting and subtle than the shadow physics
of elastic processes. In particular, inelastic diffraction is endemic, not
only in hadron-hadron collisions, but also in electron-hadron collisions,
where arguably 20 to 30 percent of all deep inelastic collisions lead to a diffractive
final state (defined as a large final-state ``rapidity gap", not
exponentially suppressed, within which no secondary hadrons are to be
found). In addition, diffractive final states are seen in a significant
fraction of hard-collision events, namely events containing
high-transverse-momentum jets, $W$s, $Z$s, and/or leptons in the final
state.

The favored descriptive tool for diffractive processes is that of the
$t$-channel exchange of a Reggeon, the so-called Pomeron. Ingelman and
Schlein, in a seminal work which created and thus far has defined the
field of hard diffraction, suggested that this Pomeron should have a
partonic description, like ordinary hadrons \cite{ref15}. This concept has greatly
helped to drive the field in a very productive manner, especially with
regard to creation of a vital and exciting experimental program.
Nevertheless, the foundations of the idea are speculative. And the field
at present is confused. The data refuse to be easily integrated into the
formalism. This was evidenced at this meeting in the excellent talk by
Hatakeyama  \cite{ref16}. I personally think that it is time to retreat from the
language of Pomeron structure functions, and to search, at the
experimental level, for more general and reliable descriptive tools to
organize and systematize the phenomenology.

In particular, it should not be taken as obvious that the $t$-channel
exchange picture is the most appropriate language. It may be that it is
better to emphasize more the $s$-channel, shadowy origins of the diffractive
phenomena, perhaps in the style of Good and Walker's original description
of diffraction dissociation \cite{ref17}, and/or of absorption models. And no matter
what, one must acknowledge that the heart of the subject resides, from a
diagrammatic point of view, in loop diagrams, not tree diagrams, and that
quantum effects, as opposed to quasiclassical partonic visualizations, are
essential no matter what descriptive viewpoint is adopted.

My favorite example for appreciating the subtlety of the phenomenon is
that of high mass, soft inelastic diffraction. A typical final-state
pseudorapidity distribution of secondary particles is illustrated in Fig.
4. First imagine that the reference frame is chosen such that zero
rapidity is at position A. Then what one would see at early times is an
ordinary multiparticle final state developing, with soft particles being
produced at large angles at early times, and more energetic particles
being produced at small angles at later times (the ``inside-outside"
cascade). But at some ``macroscopic" late time (proportional to the energy
of the fastest right-moving diffractively produced secondary), which can
in practice be tens or hundreds of fermis, the leading right-moving
emitter suddenly stops emitting: a quantum decision has been made that the
right mover (neither a nucleon or ``not a nucleon", but a quantum
superposition of each possibility) projects to the right-moving nucleon
state.

\begin{figure}[htb] 
\centerline{\epsfig{file=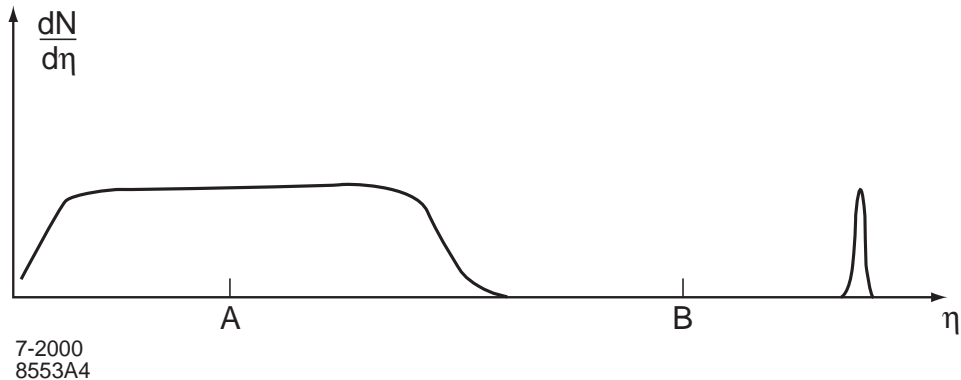,height=2in,width=5in}}
\vspace{10pt}
\caption[]{Typical pseudorapidity distribution of secondary hadrons in
massive soft inelastic diffraction.}
\label{fig4}
\end{figure}

Now view the same process in a different reference frame, where zero
pseudorapidity occurs at location B. In such a frame at early times there
is {\it no} particle emission, as if the process were at most elastic
scattering. But then, again at a ``macroscopically" late time (proportional
to the energy of the least energetic left-moving diffractive
secondary---actually the same particle as before), the left mover (which
has to be neither ``not a nucleon'' nor a nucleon, but a quantum superposition
of each possibility) makes a quantum decision {\it not} to be the nucleon
and to emit particles.

These two viewpoints represent the {\it same} physics, and a good picture
of diffraction should be able to explain how and why they are the same.  
I find this a very interesting challenge, one which the present
diagrammatic/partonic approach does not begin to touch. To do a really
good job on diffractive physics may well be the last great frontier in
hadron physics to be solved.

\subsection{Hadronization Dynamics}

A primary task of the hadron-physics subfield of multiparticle production
is to understand and describe the transition from the multipartonic
evolution at very short distance scales to the multihadronic final states
observed experimentally.  The prototypical reaction for doing this is
electron-positron annihilation into hadrons. In that case, there is a
rather satisfactory level of understanding, thanks both to the intrinsic
cleanliness of the basic process, and to the large data set of complete
events over a large energy scale.

Even so, it is interesting that two apparently competitive viewpoints,
that of the nonperturbative QCD string, {\it a la} Lund, and that of the QCD
partonic cascade, peacefully coexist. Quite sophisticated phenomena, such
as the ``string effect" in three-jet final states, can be described in
either picture with comparable success. Which is right? Conventional
wisdom seems to be that most of the space-time evolution is in fact
perturbative, with a rather quick transition to the final configuration of
emergent hadrons. (This is basically the ``preconfinement" picture).
However, were the lightest quarks to have mass of a GeV or so, then QCD
strings would not easily break, and the spacetime region of final-state
evolution would be enlarged, with an interior boundary between
perturbative evolution and stringy evolution (Fig. 5). In the
heavy quark limit the future light cone gets filled with string.  It might
be of interest to study the hadronization phenomenology as function of
quark masses in order to sort out perturbative from stringy effects.

\begin{figure}[htb] 
\centerline{\epsfig{file=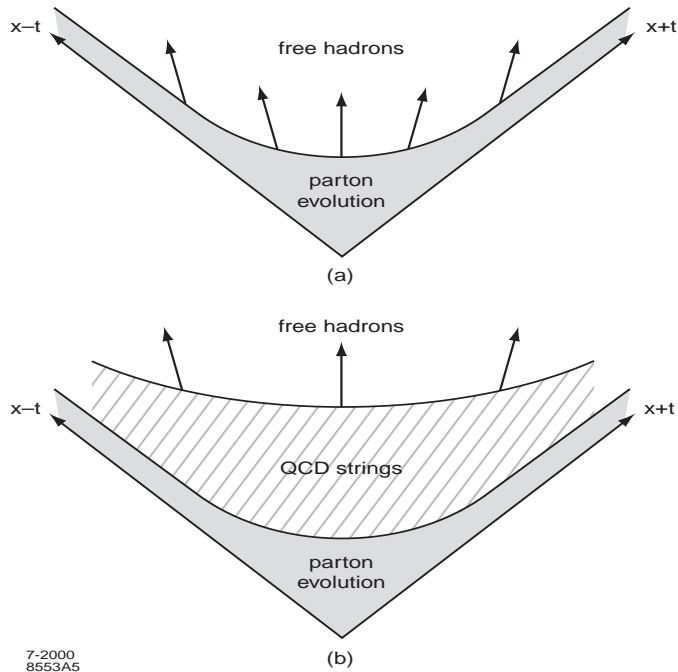,height=3.5in,width=3.5in}}
\vspace{10pt}
\caption[]{(a) Final-state evolution in dijet production in $e^+e^-$ annihilation.
(b) The same, in the case of the lightest quarks having masses $\sim$ 1
GeV.}
\label{fig5}
\end{figure}

Hadronization phenomenology in hadron-hadron collisions is different and
more difficult. The whole subject is in a much more primitive state,
especially at collider energies. Not only is the theory much harder, but
also there is a paucity of data. Much more experimental attention needs to
be paid to the generic collision phenomenology. I think it is a necessary
condition for significant progress to be made \cite{ref18}.

\section{The Future of Hadron Physics}

In the introductory section, we argued that hadron physics has social
problems that require it to be better defined and organized than at
present. The main reason that drives this notion beyond the merely
academic and that requires, in my opinion, some action is that this would
facilitate a more rational pattern of funding and support, and that it
would facilitate better access to the high-energy physics laboratories,
built and managed for purposes other than exploring the details of hadron
structure. In the previous cases of chemistry, atomic, and nuclear
physics, those who chose to specialize in those fields instead of moving
on to the higher energies and shorter distances of particle physics
could rather easily do so. Experimental facilities fit, until quite
recently, on university campuses. And the funding structures, including
peer review systems, adiabatically evolved to adapt to changes 
of scientific emphases. But hadron physics now presents itself as a
crossover field. It is beyond nuclear physics, although heavily populated
by nuclear physicists. And much of it is within the energy scale of high
energy physics, despite the fact that not many high energy physicists are
practitioners. 

Because of this, access of experimental hadron physicists to high energy
laboratories is made especially difficult.  If, in an austere fiscal
situation, a hadron physics initiative of the highest quality is put in
competition with, say, a quality next-generation neutrino experiment, there
will likely be very little support within the high-energy community for
the hadron-physics initiative. Indeed, were I myself wearing my
high-energy physics hat, I would have a hard time too. Under these
circumstances, it seems to me that the best way for hadron physics
initiatives to be viable at high energy labs is that there be independent
funding available, and that there be agreements with the high energy
laboratories for a certain amount of access to collision regions, beam lines,
luminosity, running time, infrastructure, {\it etc.} in return for 
appropriate contributions to the laboratory budgets.  Hadron physics review 
structures at the program-committee level would be essentially independent 
of those of particle physics, although at
higher policy levels there would necessarily be mixing of the communities.

Examples of this kind of setup exist. At SLAC the NPAS program allowed use
of the linac for fixed target experiments at moderate energies of interest
to the nuclear community. Despite its modest size, I have been told it was
difficult to establish. The best example seems to me to be CERN, which for
a long time has had nuclear physics as part of its program, a feature
which is now expressed in the heavy ion initiatives at the SPS and at the
LHC.  The presence of the nuclear physics component at CERN has been
important not only in providing a broader scientific base, with all the
opportunities for cross-fertilization and diversification that that
implies, but has also been useful in broadening its political base.

In the United States, Fermilab is a good example of a high energy
laboratory where hadron-physics could be pursued much more aggressively
than at present. One of the main-injector beam lines could well support
that dream next-generation multiparticle spectrometer for spectroscopy,
Regge dynamics, and correlation studies mentioned in item 4 of the
previous section. Hyperon physics and charm physics are other possible
options. Antiproton sources have been productive venues for charmonium and
other studies, both at Fermilab and CERN, and there is more to be done.
Indeed a workshop is scheduled for investigating such future options at
Fermilab \cite{refaa}. Finally,  the C0 collision region of the TeVatron collider, 
the presumed
home of the future BTeV $B$-physics initiative, is also a very attractive
venue for studying hadron physics. The topics include charm physics, low
$p_T$, diffraction, leading particle studies, and the study of collision
dynamics, especially were full-acceptance detection of complete events
available.  And the future facilities under discussion,
in particular muon colliders and/or neutrino sources based on
muon-collider technology, are rich sources of hadron-physics spinoffs.
While other spinoffs, such as $K$-decay physics, muon physics and deep
inelastic neutrino reactions, have been discussed in this context, very
little attention has been paid to the hadron-physics opportunities \cite{ref19}.

I think that a necessary condition for the situation to change is that the
hadron physics community organizes itself better. It must not only
identify itself and exhibit some political strength, but it must
also define better what hadron physics comprises,
what its fundamental scientific goals are, what the experimental programs
are that deserve the greatest attention, and what the basic challenges to
theory are.  Since organizational changes within funding agencies and
advisory structures are likely to be slow, it seems to me that the best
opportunity for getting things going might be within the professional
societies. In particular there perhaps should be a Division of Hadron
Physics within the American Physical Society. It might provide the venue
and organizational structures for achieving the above goals, and provide a
basis for going further if, as I suspect is the case, it is deemed
necessary to do so.

But most fundamental of all is that there exists a vital community of
experimental and theoretical physicists just doing hadron physics, no
matter what the obstacles. This meeting has been a splendid example that
there are at present plenty of people doing just that. We should do
everything we can to not only keep this field healthy, but to strengthen it.  
The scientific challenges will take quite some time to overcome, and in
the meantime we must make every effort to acquire the means to
overcome them.

\section{Acknowledgments}

On behalf of the participants, it is a pleasure to thank the organizers,
especially Stanley Kowalski, and Anne MacInnis, for all their hard work in
making this such a splendid meeting.


\begin{references}
\bibitem{ref18}
More detailed discussion can be found in the FELIX Letter of Intent,
Ref. \cite{ref12a}.
\bibitem{ref19}
See the talk of Shaevitz, M., these proceedings.  While one finds there some
discussion of opportunities for parton-level deep-inelastic QCD, there is 
little if any discussion of opportunities at the hadron-physics level.
\end{references}
\end{document}